# Leukemia Detection Based on Microscopic Blood Smear Images Using Deep Neural Networks

**Abdelmageed Ahmed**
dept. Engineering Electrical and Computer Engineering
University of Ottawa
Cairo, Egypt
ahass202@uottawa.ca

**Alaa Nagy**
dept. Engineering Electrical and Computer Engineering
University of Ottawa
Cairo, Egypt
aelba046@uottawa.ca

**Ahmed Kamal**
*dept. biomedical engineering department*
Minai university
Minya, Egypt
ahmd654@yahoo.com

**Daila Farghl**
*dept. biomedical engineering department*
Minai university
Minya, Egypt
dolly.mostafa93@yahoo.com

*Abstract*— In this paper we discuss a new method for detecting leukemia in microscopic blood smear images using deep neural networks to diagnose leukemia early in blood. leukemia is considered one of the most dangerous mortality causes for a human being, the traditional process of diagnosis of leukemia in blood is complex, costly, and time-consuming, so patients could not receive medical treatment on time; Computer vision classification technique using deep learning can overcome the problems of traditional analysis of blood smears, our system for leukemia detection provides 97.3 % accuracy in classifying samples as cancerous or normal samples by taking a shot of blood smear and passing it as an input to the system that will check whether it contains cancer or not. In case of containing cancer cells, then the hematological expert passes the sample to a more complex device such as flow cytometry to generate complete information about the progress of cancer in the blood.

## I. INTRODUCTION

Leukemia is a type of cancer affecting blood; if it is detected late, it will result in death. Leukemia develops when the bone marrow produces an excessive number of aberrant white blood cells. The normal of the blood system will be disrupted when aberrant white blood cells are in excess. Hematologists can identify abnormal blood when they draw a blood sample and study it[1]. However, hematologists will inspect microscopic images visually, and the process is time-consuming and tiring [1 - 3]. Moreover, the process requires human experts and is prone to errors due to emotional disturbance and human physical capability, which has its limitations.

Moreover, it is not easy to get consistent results from visual inspection [1]. Visual inspection can only give qualitative results for further research [1]. Studies indicate that the majority of modern methods. Use all blood-related data, such as the number of red blood cells, hemoglobin level, hematocrit level, mean corpuscular volume, and much more, as the criterion for categorizing disorders like cancer, thalassemia, Etc. Expensive testing and equipment labs are required to know all information about blood. An automatic image processing system is urgently needed and can overcome related constraints in visual inspection. The system to be developed will be based on microscopic images to recognize leukemia cells in blood smears. The early and fast identification of the leukemia type greatly aids in providing the appropriate treatment for a particular type of leukemia [4]. The currently used diagnostic methods rely on analyzing immuno- phenotyping, fluorescence in situ hybridization, cytogenetic analysis, and cytochemistry. Sophisticated and expensive laboratories are required in order to run the diagnostic methods, and it has been reported to provide a high ratio of misidentification; with this system, more images can be processed, reduce analyzing time, exclude the influence of subjective factors, and increase the accuracy of identification process at the same time. In machine learning, the inspection and classification of leukemia will be based on the texture, shape, size, color, and statistical analysis of white blood cells.

In contrast, deep learning makes it much more profound and gets the whole image's exclusive features. This project is applied to increase efficiency globally and can simultaneously benefit and be a massive contribution to the medical and pattern recognition field. The main objective is to enhance algorithms that can extract data from human blood where human blood is the primary source to detect diseases at an earlier stage and can prevent it quickly [5]. This system should be robust towards diversity among individuals, sample collection protocols, time, Etc. This automated system can produce lab results quickly, easily, and efficiently.

## II. DATASET

Images that were used in this project were downloaded from the internet and are available in ALL IDB[6], ASH Image Bank Hematology [7], Stock photo, vectors and Royalty-free Images[8], Shutter stock[9], Atlas of Hematology [10], Atlas of blood smear analysis[11], Blue Histology and American Society of Hematology [12], This dataset is composed of 630 images, contains 480 cancer images and 150 normal images.

## III. METHODOLOGY

### A. Data Preprocessing

#### 1) Remove duplication

As the dataset is collected from various resources, had found that there are some repetitions, some images contain a watermark, and other contains websites' logo totally about 43 images, so now the data set has become 587 images.

#### 2) Resizing of images

As the dataset has a different distribution of size, and for training the CNN model, it was needed to make all images in the dataset has the same size, so we applied a resizing technique and make all image 256 x 256 pixels to reduce the training time. as shown in figure [5.1]

#### 3) Filtering images

Before the processing stage, we need to remove noise and enhance line structures in images [13], and this is available by applying a median filter (3 x3) and sharpening the image (3 x3) ,as shown in Fig[1].

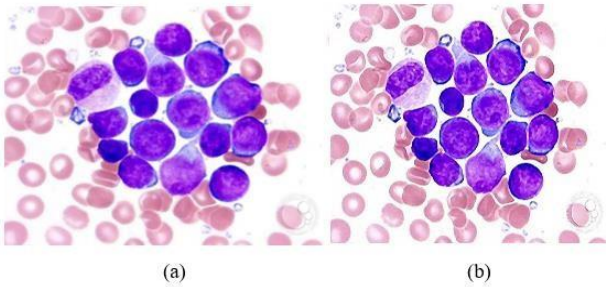

Figure 1:(a) original image,(b) image resized by 256*256 and filtered by median and sharpen filters

#### 4) Data augmentation

Image data augmentation is a method for artificially increasing the size of a training dataset by producing altered copies of the dataset's images [14]. The capacity of fit models to generalize what they have learned to new pictures may be improved by training deep-learning neural network models on more data. Additionally, augmentation techniques can provide variants of the images. Through the ImageDataGenerator class, the Keras deep learning neural network framework can fit models by adding picture data [15]. There are many different types of augmentation techniques, some of them as:

##### a) Flipping

An image flip means reversing the rows or columns of pixels in the case of a vertical or horizontal flip [9].

##### b) Horizontal and Vertical Shift Augmentation

A shift to an image means moving all pixels of the image in one direction, such as horizontally or vertically, while keeping the image dimensions the same; this means that some of the pixels will be clipped off the image, and there will be a region of the image where new pixel values will have to be specified [16].

##### c) Random Zoom Augmentation

A zoom augmentation randomly enlarges the image and either interpolate or adds new pixel values around the image [16].

##### d) Shearing

Shearing will automatically crop the correct area from the sheared image so that we have an image with no black space or padding [16].

##### e) Interpolation (Nearest)

A technique for creating new data points within the range of a discrete set of existing data points is interpolation [17]. Nearest neighbor interpolation is the most straightforward approach to interpolation. Rather than calculate an average value by some weighting criteria or generate an intermediate value based on complicated rules, this method simply determines the "nearest" neighboring pixel and assumes its intensity value of it [13]. And Fig[2] indicates a sample image with its augmented one.

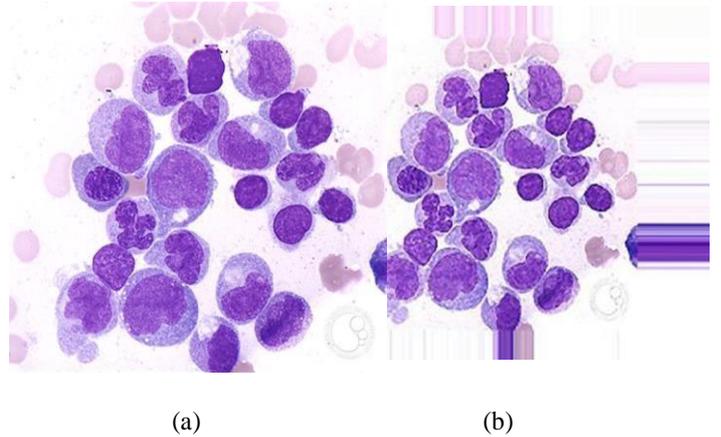

Fig 2: (a) original image and (b) augmented image.

### B. Processing stage

After augmentation processes, our data become 1550 images for cancer and 1480 for normal. To fit data to models, we divided it through coding into three data sets: training set, validation set, and test set by ratios 60%, 20%, and 20%, respectively. Then the next stage is to train the model that can be able to classify the images.

Our optimizing parameters are accuracy and validation accuracy: to get the best of them as possible, we trained three networks with different architectures.

##### a) BasicCNN model

In this model, the input images were (RGB) color images with a resolution of 128x128 pixels. It consists of 3 convolutional layers with max pooling layers. A rectified linear unit follows each convolutional layer (relu). We used a constant filter size (3x3), and the number of

Filters (128), the stride of ones (equal 1), and fully connected layers trained for two categories classification using the sigmoid activation function. Where we classified the data set into leukemia cells or normal cells, this architect achieved 90.99% accuracy and 84.97 % validation accuracy after 17 epochs.

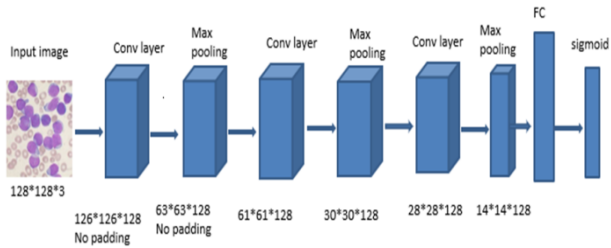

Fig.3: Indicates the block diagram of the basic CNN model

*b) Alexnet architecture*

In this study, we deployed the pre-trained AlexNet to detect ALL and classify its subtypes. This architecture was proposed by Krizhevsky et al., nine who deployed this architecture for the ImageNet Large Scale Visual Recognition Challenge 2012,20 and won the challenge in the first place. Input images were Red, Green, and Blue (RGB) color images with a resolution of 227 x 227 pixels. It consists of 5 convolutional layers with three max polling layers. Each convolutional layer in AlexNet architecture is followed by a rectified linear unit (ReLU). All the parameters, including the filter size, the number of filters, and the stride for each layer, are illustrated in Fig.4; we replaced the SoftMax layer with a sigmoid layer as we want to classify the input image into only two types of this architect achieved 55.35% accuracy and 49.76 % validation accuracy after 12 epochs.

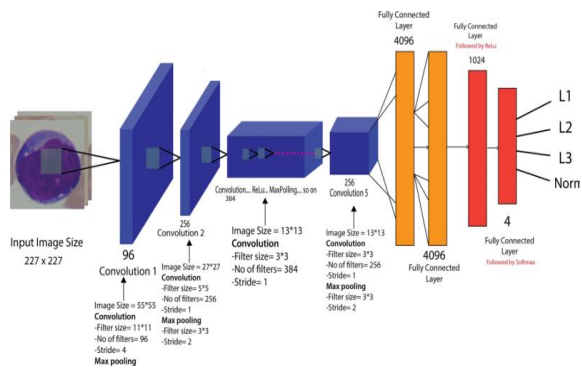

Figure 4: AlexNet architecture for acute lymphoblastic leukemia subtype classification. Last 2 layers are newly added.

*c) Modification of model used in published paper*

This used a retrained model that had been used in a published paper [20], shown in figure 5, and we changed the values of the hyperparameter to become as shown in figure 6; This network contains five layers. The first three layers perform feature extraction, and the other two layers (fully connected and SoftMax) classify the extracted features. The input image has a size of 128x128x3. In convolution layer 1, we used a constant filter size of 5x5 and a total of 16 different filters. The stride is one, and no zero-padding was applied. The second and third convolution layers have the same structure as the first one but a different number of filters, 32 and 64, respectively. We used a pooling layer with filter size two and stride 2 to decrease the volume spatially. During the model we learned, the mini-batch's chosen size was 128, and ReLu was used as the activation function. This architect gives: accuracy = 97.73 % validation accuracy = 94.64 %

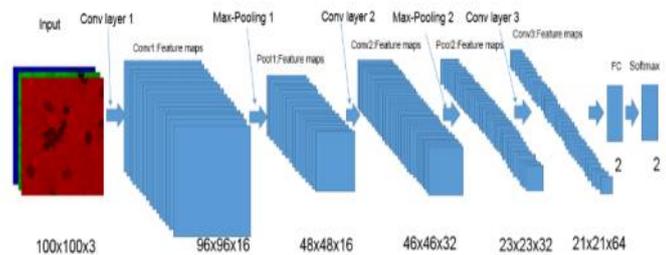

Fig. 5: The original architecture of CNN in the mentioned paper.

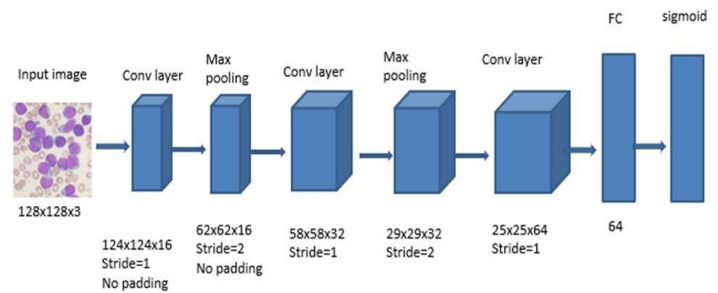

Fig. 6: Architecture of CNN after changes in hyperparameter

## IV. EXPREMENTL RESULT

Our experiments were conducted on Python 3.7 with 3030 images, 60% (1818 images) of them for training, 20% (606 images) for validation, and the remaining 20% (606 images) for testing our model. In order to evaluate each model and clarify the best one, we compare them by some statistically measured parameters:

*A. Accuracy*

**Train accuracy**

For the basic CNN model, train accuracy comes to 90.99% after 17 epochs; our leukemia classifier is doing an excellent classification, as shown in fig.7.1a. For AlexNet architecture, the accuracy achieved its maximum accuracy of 56% after 11 epochs; that means our model is terrible on leukemia classification as shown in fig.6.1 b, but the Modification of the model used in Thanh et al. paper [18] achieved the maximum accuracy over all models 97.73 % after ten epochs as shown in fig.6.1c. 6.1.3

**Validation accuracy**

Basic CNN Model validation accuracy reaches 85% after 17 epochs, as shown in fig.7.1a. Therefore, we expect our model to perform with ~85% accuracy on new data. For AlexNet architecture, the accuracy achieved its maximum accuracy of 53.6% after 11 epochs; that means our model is terrible on leukemia classification, as shown in fig.6.1 b; This means that we expect our model to perform with ~53.6% accuracy on new data. Nevertheless, in Modification of the model used in Thanh et al. paper [94] achieved the maximum validation accuracy over all models at 94.3 % after ten epochs, as shown in fig.6.1c. Therefore, we expect our model to perform with ~94.3 % accuracy on new data. We notice that our train metric increases as epochs increase while the validation accuracy metric decreases. That means that our model fits the training set better but slightly loses its ability to predict new data, indicating that our models are beginning to overfit.

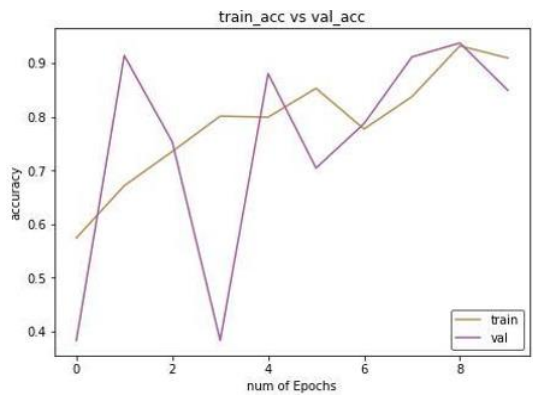

Fig.7.1a, curve of val acc & train acc for basic CNN model

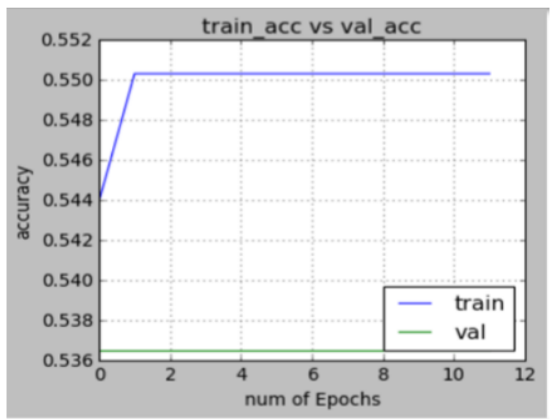

Fig.7.2b, curve of val acc & train acc for AlexNet architecture

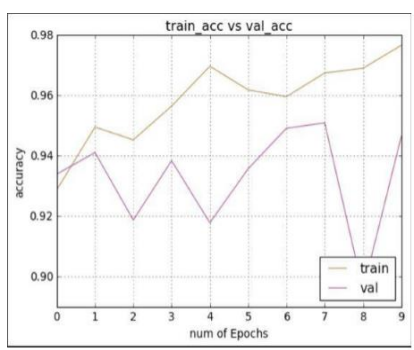

Fig.7.2c, curve of validation accuracy & train accuracy for Modification of model used in Thanh et al paper [18]

B. *Confusion Matrix*

A classification problem's predicted outcomes are compiled in a confusion matrix. The count values describe the number of accurate and inaccurate predictions for each class. Because it is feasible to see the relationships between the classifier outputs and the real ones, this is a great alternative for reporting results in M-class classification issues. For the basic CNN model, the number of leukemia images that are predicted as leukemia is 372, the number of leukemia images that are predicted as normal is 8, the number of normal images predicted as normal is 269, and the number of normal images that are predicted as leukemia is 51, as shown in fig. 7.3 a. These accuracies show that this model is good at predicting leukemia images but bad at predicting normal images. For AlexNet architecture, the number of leukemia images that are predicted as leukemia is 0, the number of leukemia images that are predicted as normal is 380, the number of normal images predicted as normal is 157, and the number of normal images that are predicted as leukemia is 163, as shown in fig.7.3 b. These accuracies show that this model is terrible at predicting normal images. The number of leukemia images predicted as leukemia for the modified model used in the published paper [18] is 369, the number of leukemia images predicted as normal is 11, the number of normal images predicted as normal is 301, and the number of normal images predicted as leukemia is 19; as shown in fig.7.3 c. These accuracies show that this model has done a great job of predicting normal images.

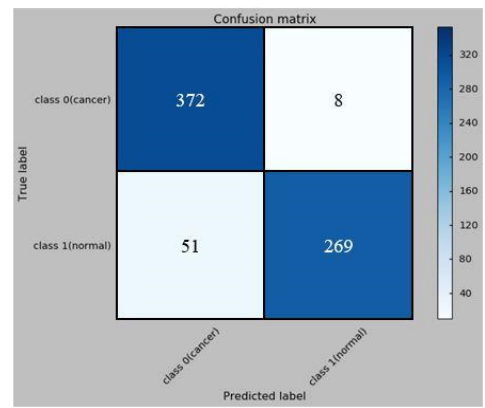

Fig.7.3a, Confusion matrix of basic CNN model

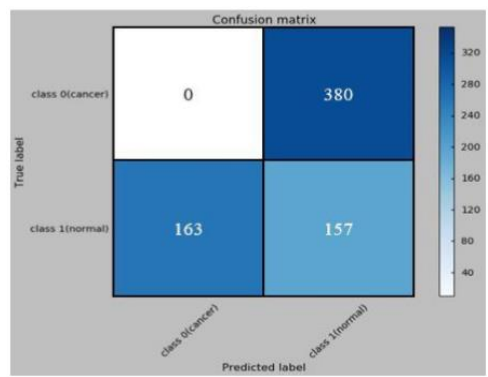

Fig.7.3b, Confusion matrix of AlexNet Architecture

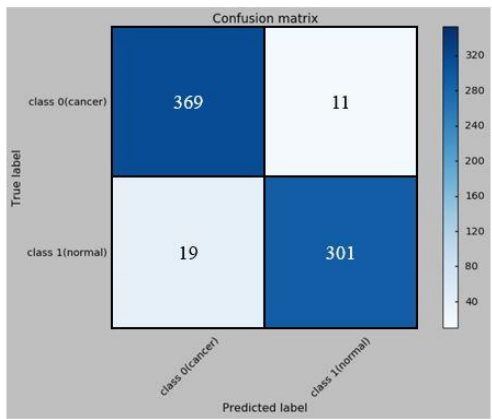

Fig.7.3c, Confusion matrix for Modification of model used in Thanh et al paper [18]

*C. Percsision*

It is calculated as the proportion of accurate positive results to those that the classifier predicted to be positive. Our CNN model has medium precision, AlexNet architecture has very low precision, and the modified version of the model used in Thanh et al.'s [18] paper has good precision due to its goodness method, as shown in fig. 7.4a.

*D. Recall*

It is determined by dividing the total number of pertinent samples (all samples that should have been labeled as positive) by the total number of reliable positive results. As illustrated in fig. 7.4a for our CNN model, fig. 7.4b for the AlexNet architecture, and fig. 7.4c for the Thanh et al. paper [18]. The perfect model regarded recall is the third model.
The first CNN model in class 1 has a high recall but low precision. This means that most of the positive examples are correctly recognized (low FN), but there are a lot of false positives. Nevertheless, in class 0, low recall and high precision show that we miss a lot of positive examples (high FN), but those we predict as positive are indeed positive (low FP).

*E. F1 Score*

The harmonic mean of recall and accuracy is the F1 score. The F1 score has a range of [0, 1]. It tells how accurate the classifier is (how many instances it classifies correctly) and how robust it is (it recognizes a significant number of instances). As illustrated in fig. 7.4a for our CNN model, fig. 7.4b for the AlexNet architecture, and fig. 7.4c for the Thanh et al. paper [18]. These figures show that the modification of the model used in Thanh et al.'s paper [18] is precise and robust.

*F. Support*

Support is the number of samples accurately representing the response within that category.
It provides information on the precise numbers of each class in the test data.
Figures 7.4a and 7.4b for the fundamental CNN model, 7.4b for the AlexNet architecture, and 6.8c for a modified version of the model from the Thanh et al. work [18] serve as examples.

```
                    precision  recall  f1-score  support
class 0(Cancerous)     0.98      0.62    0.76       78
class 1(Normal)        0.84      0.99    0.91      163
```

Fig.7.4a, values of precision, recall, f1 score and support for our CNN model

```
                precision  recall  f1-score  support
class 1(cancer)    0.00     0.00    0.00      178
class 0(normal)    0.49     1.00    0.66      172
```

Fig.7.4b, values of precision, recall, f1 score and support for AlexNet architecture

```
                precision  recall  f1-score  support
class 0(cancer)    0.97     0.95    0.96      373
class 1(normal)    0.94     0.97    0.95      327
```

Fig.7.4c, values of precision, recall, f1 score and support for Modification of model used in Thanh et al paper [18]

## V. DISCUSSION

Leukemia is a malignancy that affects the body's blood-forming tissues, including the lymphatic system and bone marrow. To get the most effective treatment, the patient needs early Diagnosis, so we deploy three models using the power of CNN to classify blood smears into normal and abnormal.
Our dataset had not been taken under the same conditions as it was collected from various resources, and it needed to be bigger to use with DL. To overcome this problem, we used the power of data augmentation; this solution was suitable for us; our data before augmentation was 260 images, and after augmentation became 3030 images.
Our optimizing parameters were accuracy and validation accuracy; by using CNN, we trained the model:

- the First model consists of 3 convolutional layers with max pooling layers. Its accuracy was 90% and 84.97 % validation accuracy. It was terrible with our dataset due to its few layers, so we trained another model
- the Second model was AlexNet; this architecture proved its efficiency in CNN models, so we trained it with our data, input is (RGB) color images with a resolution of 227 x 227 pixels. It consists of 5 convolutional layers with three max polling layers. These models achieved 55.35% accuracy and 49.76 % validation accuracy. We found that it does not fit our dataset. So we still have the same problem of low accuracy and keep looking for another model
- In the last model, we used a retrained model that had been used in a published paper [18]; it contain7 layers. The first five layers perform feature extraction, and the other two layers (fully connected and SoftMax) classify the extracted features. The input image has a size of 128x128x3. This architect has an accuracy of 97.73 % validation accuracy is 94.64 %, finally, we found that this model fit our data

- 

## VI. CONCLUSIONS

In this system, we investigated the application of deep CNNs. We deployed a pre-trained model for detecting and classifying the blood sample into normal and abnormal samples using microscopic blood sample images and convolutional neural network classification algorithms. The system was built by deep learning, which uses all features in microscopic images, not only examining changes of specific features as a classifier input. We have performed the pre-trained model in a largely augmented dataset to confirm the system's accuracy and reliability. By performing data augmentation, we can achieve 97.3% accuracy. The system has high accuracy, and less processing time (show results in less than 30 seconds)
, minor errors, and early identification of leukemia successful in giving the patient the proper care. And cheaper cost.

The detection system was built in three parts:
1) the acquisition part, which consists of a digital camera that has been installed at the top of the eyepiece of the microscope,
2) pre-trained CNN model responsible for the classification system.
3) a graphical user interface to display the image obtained from the camera and show the classification results.

## VII. FUTURE WORK

Expanding the focus on classifying the subtypes of leukemia cells such as Acute Myeloid Leukemia or AML, Chronic Myeloid Leukemia or CML, Acute Lymphoid Leukemia or ALL, and Chronic Lymphoid Leukemia or CLL not only separating between cancerous and non-cancerous cells and developing a convenient environment to construct an extensive leukemia dataset as this topic of research suffer from leaks in images.